# Tunable Optical Bistability and Optical Switching by Nonlinear Metamaterials

Sinhara R Silva, Alexander D Shields, Jiangfeng Zhou

*University of South Florida*

**Abstract**

We demonstrate a nonlinear metamaterial in microwave frequency regime with hysteresis effect and bistable states, which can be utilized as a remotely controllable micro second switching device. A varactor loaded split-ring resonator (SRR) design which exhibits power and frequency dependent broadband tunability of the resonance frequency for an external control signal is used. More importantly, the SRR shows bistability with distinct transmission levels. The transition between bi-states is controlled by impulses of an external pump signal. Furthermore, we experimentally demonstrate that transition rate is in the order of microseconds by using a varactor loaded double split-ring resonator (DSRR) design composed of two concentric rings.

**Introduction**

Electromagnetic Metamaterials are artificial structures periodically arranged to exhibit fascinating electromagnetic properties, not existing in nature [1]. One example is a synthetic homogeneous media that consists of element structures termed meta-atoms which are engineered to have both a negative permittivity and a negative permeability [2-4]. These structures introduce peculiar behaviors such as negative index of refraction [5, 6], super-resolution or perfect lensing [7], and electromagnetic cloaking [8, 9].

A great deal of research in the field of metamaterial was conducted in a linear regime, where the electromagnetic responses are independent of the external electric or magnetic fields. Unfortunately, in the linear regime the desired properties of metamaterials have only been achieved within a narrow bandwidth around a fixed frequency. There are a few ways that passive tunability in metamaterials can be achieved. Some methods include geometric modification, insertion of variable resistors, and insertion of variable capacitors. The introduction of additional controllable media [14-16] such as nonlinear elements to the structure has led to extensive studies on nonlinear metamaterials [16-18]. In turn, nonlinear metamaterials have given way to broadband resonance responses in metamaterials. Amongst various nonlinear elements, introducing a varactor diode to the structure has been the most preferred and more effective method for obtaining broadband responses in the microwave and optical regime. It has been demonstrated that applying a DC bias voltage [19-21], externally applied electromagnetic fields [15, 22, 23], and even heat [10, 24] can be used to achieve non-linearity and tunability in varactor loaded metamaterials.

Among basic metamaterial structures split ring resonators are known as the most common and best characterized [15, 25] metamaterial with geometrically scalable meta-atoms which can be translated to operability in many decades of frequencies. In this paper, we present a nonlinear tunable metamaterial operating at microwave frequencies. We demonstrate that an external pump signal can be used to remotely tune the SRR resonance frequency. Furthermore, we investigate the bistability and hysteresis behavior of the SRR under treatment of an external pump. We also show that the external pump can be used to remotely trigger the SRR between bistable resonance states. The SRR sample has nonlinear behavior that was achieved by mounting a hyper-abrupt tuning varactor diode [26, 27] as a bridge for one of the gaps in the SRR.

**Design and the experimental set up**

The resonator structures are fabricated using copper on a FR4 substrate of thickness 0.2mm. A PNA-X Network Analyzer is used to measure the S-parameters of the SRR sample while a N5181A MXG Analog Signal Generator is used as the pump signal. The experiments were performed with both the pump and the probe signals in such a way as to have the electric field polarization parallel to the vacant gap of the SRR structure.

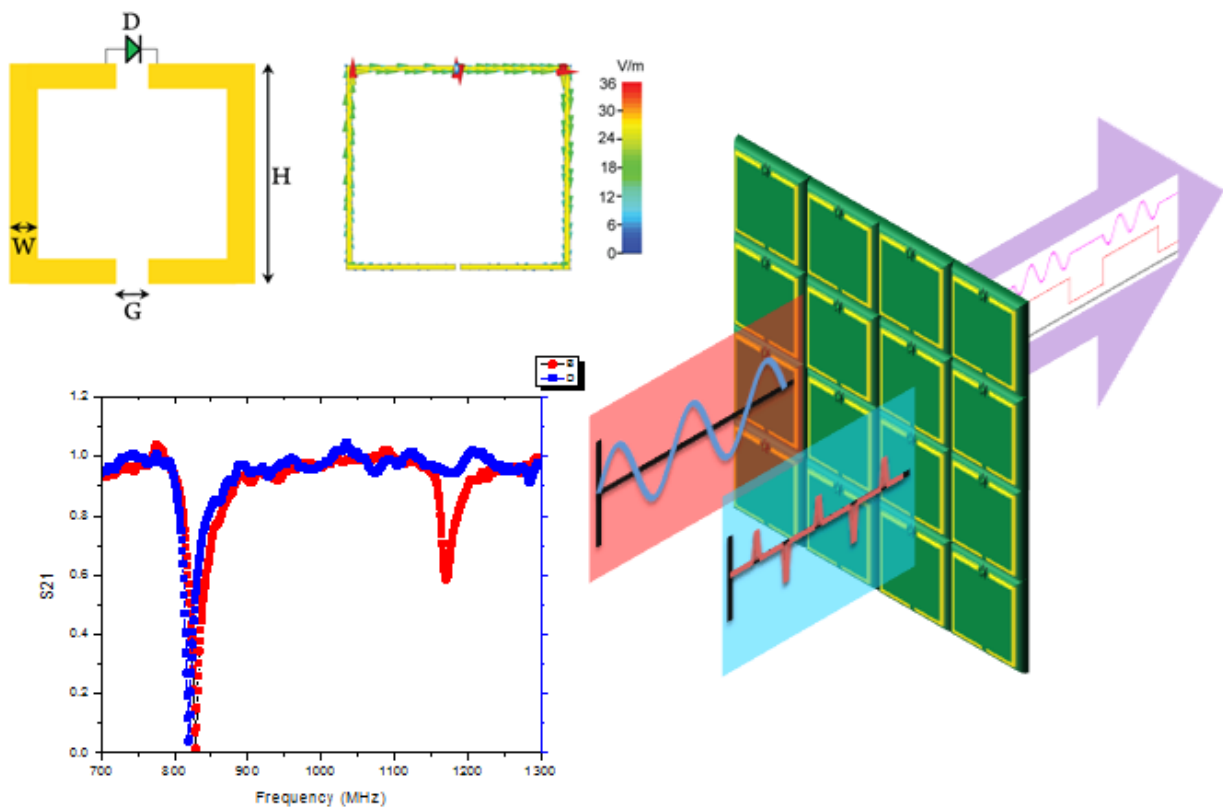

**Figure 1: Experimental design of the SRR structure**
**a)** Experimental configuration of the metamaterial behavior as a switching device **b)** Geometry and dimensions of the SRR structure. H=40mm, h=20mm, G=0.5mm, W=1mm, where (D) is a Hyper abrupt varactor diode **c)** Surface current density obtained by finite element simulations using CST.

Figure 1(a) illustrates the SRR being controlled by external signals and being used as a switchable device in the microwave frequency regime. The SRR structure was exposed to an external signal for both increasing and decreasing pump frequency with a constant power as well as both increasing and decreasing pump power at constant frequencies. The behavior of the SRR's transmission was observed using the PNA-X Network Analyzer.

**Hysteresis and Bistability**

Using the PNA-X Network Analyzer, S-parameters of the sample are measured under varying conditions. To clearly show bistability as well as a hysteresis effect, measurements of the SRR under increasing pump frequency and decreasing pump frequency were compared as shown in fig 2.

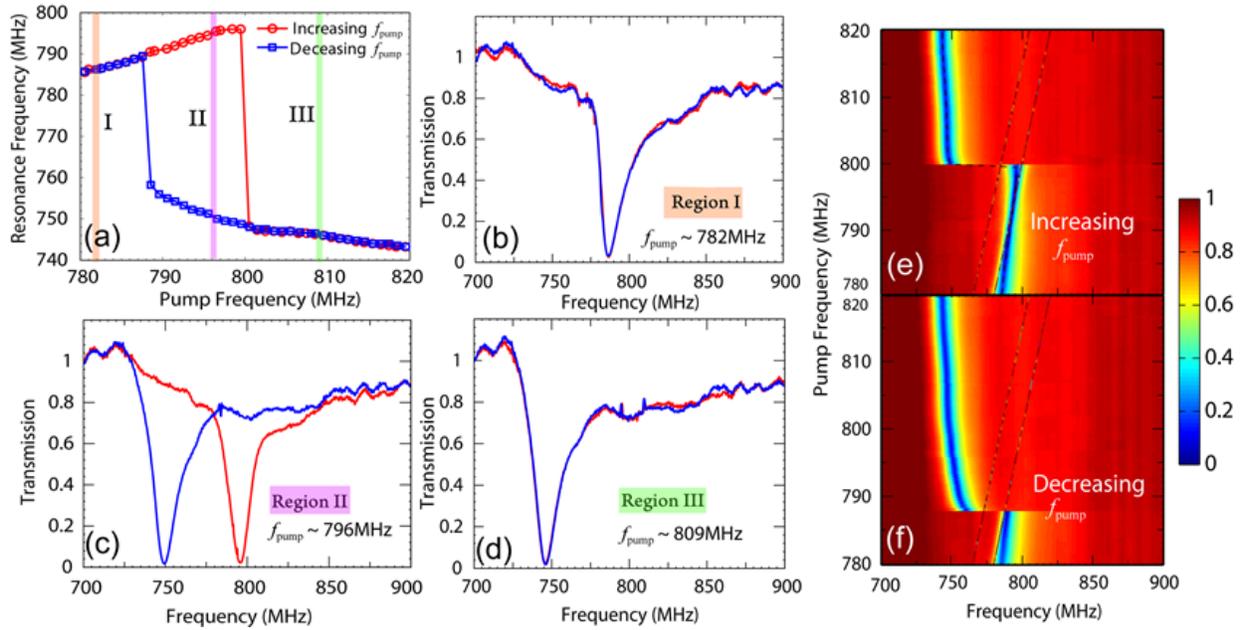

**Figure 2. Resonance frequency dependence for continuously increasing (red) and decreasing (blue) external signal frequency with constant power P*pump*=-25dBm**

**a)** Hysteresis behavior of the frequency dependent resonance frequency  **b), c), d),** Frequency dependant transmitted intensity for pump frequency *fpump* = 782MHz(region 1), *fpump* = 796MHz(region 2) , *fpump* = 809MHz(region 3), respectively **e), f),** frequency dependent 2D transmitted intensity behavior.

The Varactor loaded SRR structure displays a dynamic tuning range of 60MHz for an external pump signal of a constant power of -25dBm. As shown in Figure 2(a), the width of the hysteresis loop is more than 20MHz for that particular intensity. Figures 2(b), 2(c), and 2(d) display the transmission behavior for the corresponding regions indicated in figure 2(a). In Regions 1 and 3 the frequency dependent transmitted intensity for decreasing and increasing pump frequencies are similar. But in region 2, it follows two different paths demonstrating bi-stability of the metamaterial structure. We observed the sudden jump between two resonance frequencies when the external pump frequency coincides with the resonance frequency of the structure.

The hysteresis-type behavior and bistability[26] of the resonance frequency can be seen not only for the external signal frequency as shown in figure 3(a), but also for external signal intensities with a constant frequency as shown in figure 3(b).

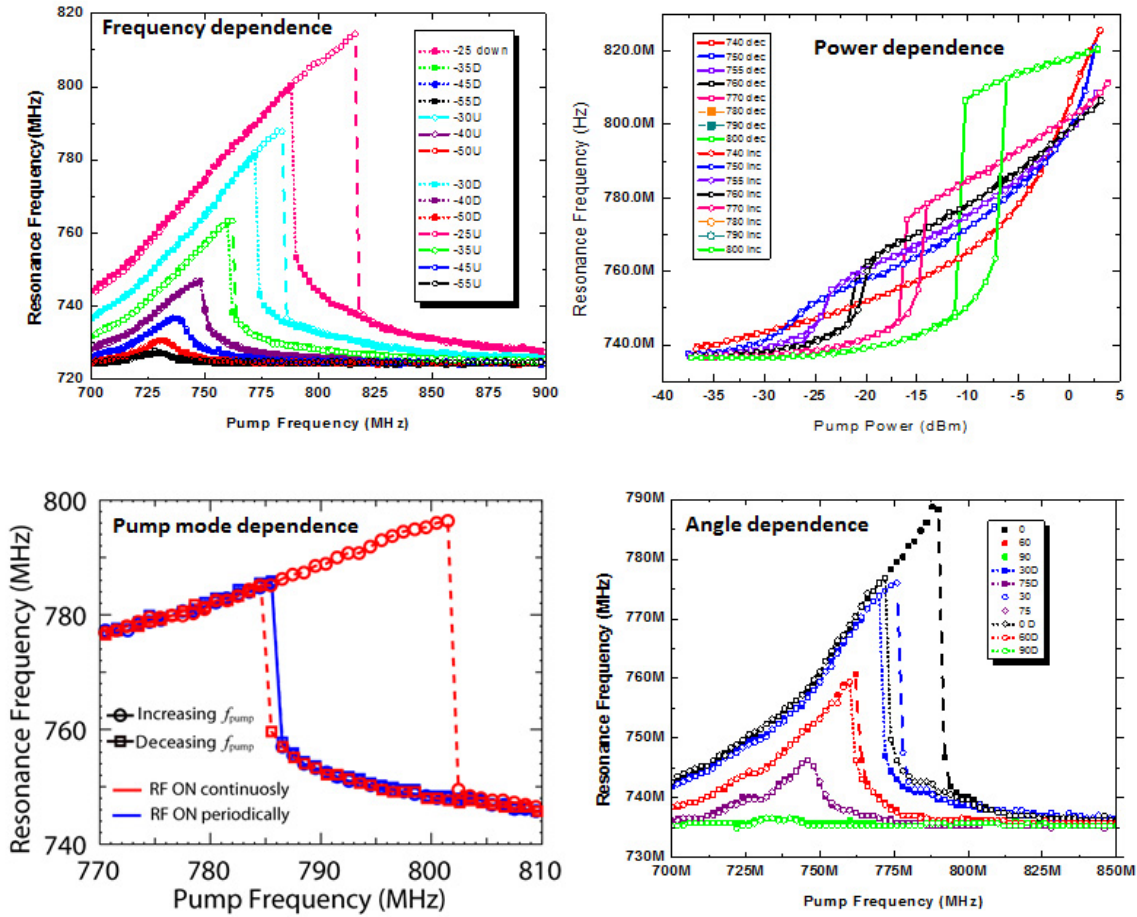

**Figure 3. Resonance frequency dependence for external signal**

Resonance frequency dependence for continuously increasing and decreasing **a)** Pump frequency **b)** Pump power **c)** Resonance frequency dependence for continuously and periodically changing pump frequency at P*pump* = -25dBm **d)** resonance frequency dependence with the angle of the E filed (Black : E field along the gap of the structure, Green : E field is perpendicular to the structure)

For very low pump powers and frequencies, hysteresis behavior vanishes and we are no longer able to observe the jump between two states. In addition, the width of the hysteresis loop widens and narrows with the increase and decrease of both the external pump signal power and frequency indicating an increase in nonlinearity. Furthermore, the direction of the sudden jump in the resonance frequency is different for varying power and varying frequency. When increasing pump power, the resonance frequency jumps from a lower frequency to a higher frequency while the opposite is true for increasing the pump frequency. We found success when modeling our structure as an RLC circuit. With the increase of the pump power, the voltage across the diode increases, which results in a decrease in the total capacitance of the structure and an increase in the resonance frequency. But, sudden transition of the resonance frequency for external frequency dependence, behaves opposite to external signal intensity dependence. Furthermore, in order to observe the hysteresis effect external pump should be applied continuously as shown in fig 3c otherwise jump is observed but not hysteresis effect. Polarization angle of the E filed with respect to the gap of the SRR has an impact on the behavior of the structure as shown in fig 3d. When the E filed is parallel to the gap of the structure we observe the bistable states but with the change of the angle for the same power level the magnitude of the e filed component along the gap reduces and shrink the hysteresis gap size.

**Switching device and transition time**

The SRR structure is improved by inserting a 1M resistor in parallel to the nonlinear varactor diode. This modification of the structure decreases the transition time between the states and improves its applicability, but minimum step size of the time was limited by the delay time of the instruments used in the experiment. Furthermore, by using a pulse train of appropriate powers, the hysteresis in figure 4(d) was observed which clearly indicated two states in the transmission.

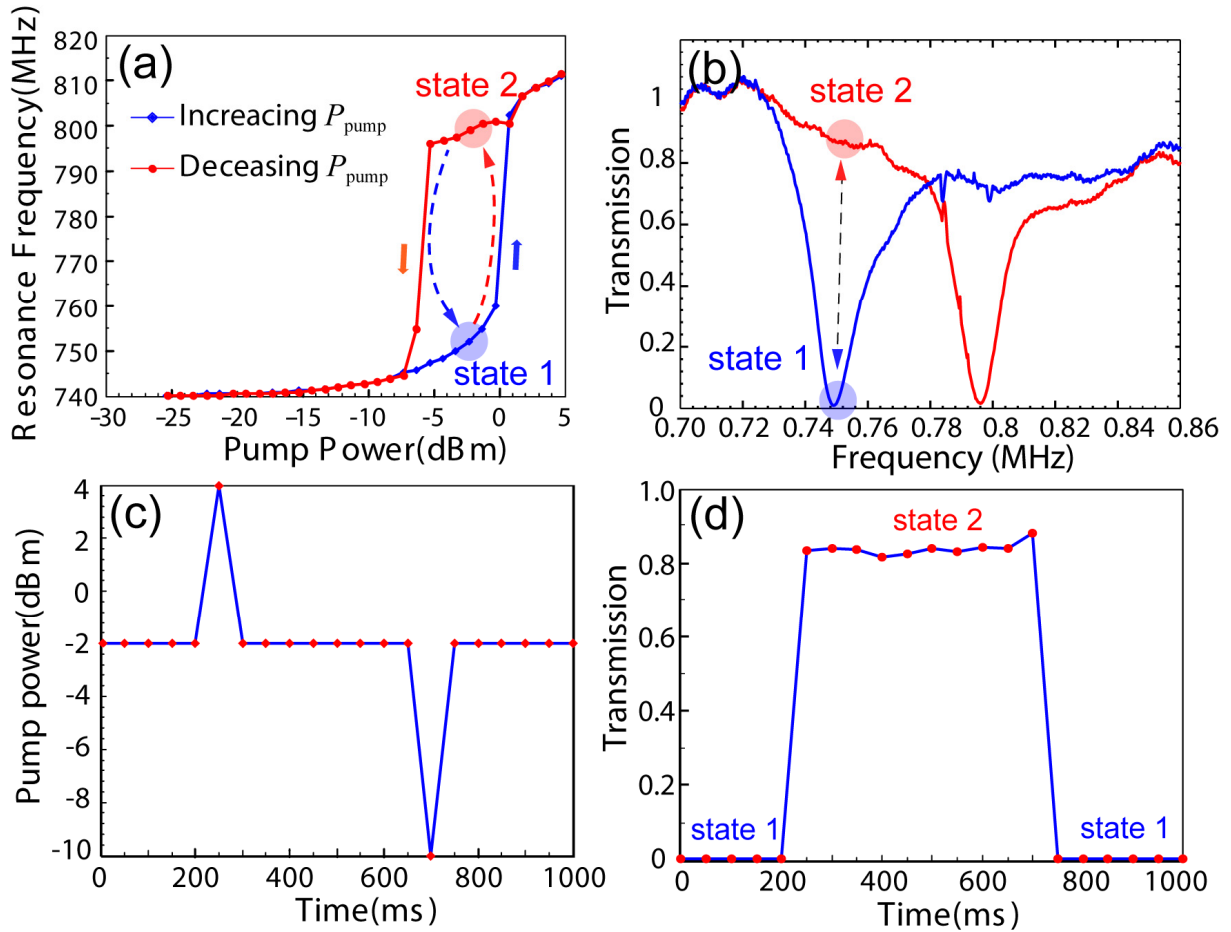

**Figure 4. Trigger transitions between the two states**

**a)** Transition between states for power dependent transmitted intensity **b)** S parameter plots of the two states **c)** Trigger signal **d)** Bistable states of transmission signal

To explore the transition of the sample under a greater resolution, different measurement techniques must be utilized. In order to capture the transition, we designed a double split ring resonator (DSRR) that could be pumped at the resonant frequency (1200MHz) of inner ring and probed at the frequency of outer ring. The power of the pump signal is about 10 dBm greater in magnitude than the signal used to probe. Figure 5a shows the new design which was used to capture the transition times in microsecond speed.

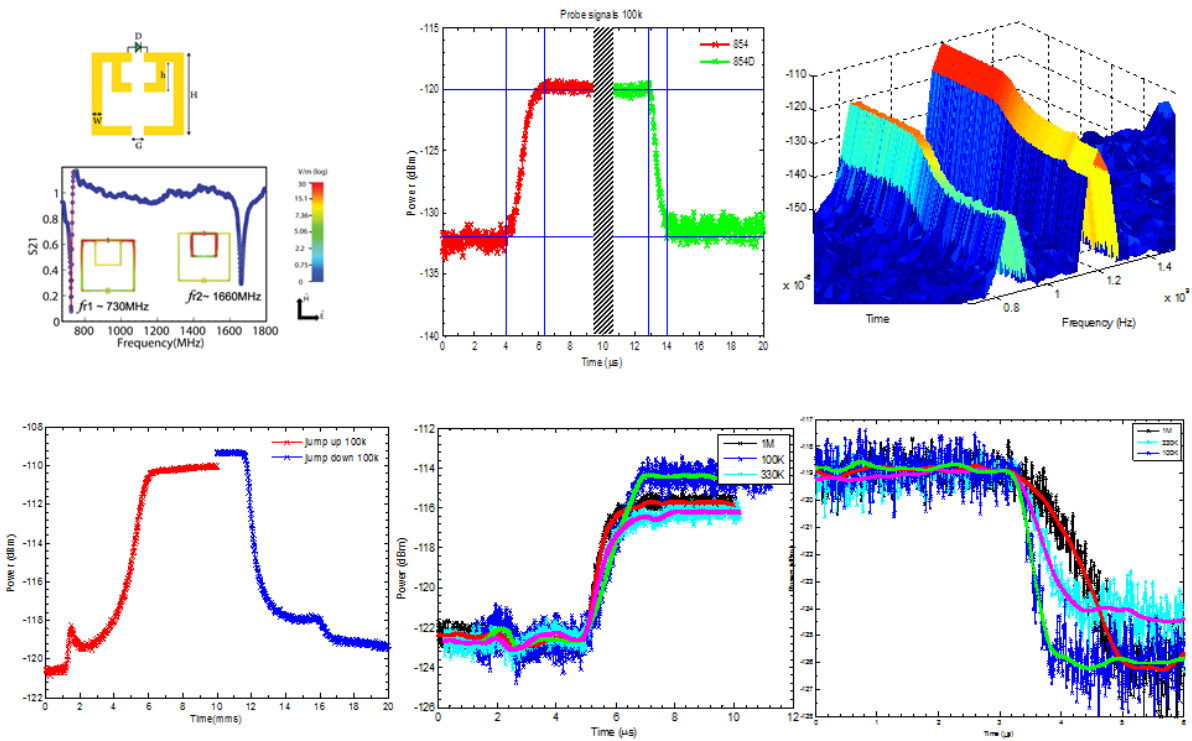

**Figure 5. a) DSRR structure and S parameter plot b) transition between two states c) spectrogram of a transition d) Trigger signal of from the pump e) and f) transition time dependence with the resistance**

As shown in fig 5b by inserting a resistor with parallel to the varactor diode of the structure we can increase the transition time by allowing a path to discharge the accumulated charge across the diode. We experimentally demonstrate that by choosing a particular resistor value jump up transition and jump down transition time are different (Fig 5b) and the transition time depends on the resistor value (fig 5e and fig 5f). In order to make a switching device with a switching speed in microseconds we need to choose a correct resistor which optimizes the both up and down transition times.

Currently, metamaterials with frequency tunability are particular of interest due to the flexibility of frequency control and has become an essential part of metamaterial devices. It makes material devices more versatile, adaptive to varying external disturbances and changes its effective parameters accordingly. Nonlinear metamaterials provide the basis for manipulating and controlling of the characteristics of a media. Therefore, the functionality demonstrated above will benefit to better understanding of the nonlinear behavior of the metamaterials. Furthermore, difference between power dependent and frequency dependent hysteresis behavior, its increased tunability and usefulness of bistable states as a switchable device in microwave. Finally,